\documentclass[a4paper,11pt]{article}
\pdfoutput=1
\usepackage{jheppub_kim}

\usepackage{graphicx}
\usepackage{amsmath}
\usepackage{amssymb}
\usepackage{amsfonts}
\usepackage{mathrsfs}
\usepackage{lscape}
\usepackage{hyperref}
\usepackage{verbatim}
\usepackage{xcolor}
\usepackage{bm}
\UseRawInputEncoding
\usepackage{tabularx}
\usepackage{mathtools}
\usepackage{ulem}
\makeatother
\title{A coupled generalized three-form dark energy model}

\author[\star]{Yan-Hong Yao,\note{Corresponding author.}}
\author[\dagger]{Xin-He Meng,}

\affiliation{Department of Physics, Nankai University, Tianjin 300071, China}

\emailAdd{$\star$ yhy@mail.nankai.edu.cn}
\emailAdd{\mbox{$\dagger$ xhm@nankai.edu.cn}}

\date{\today}

\abstract{A coupled dark energy model is considered, in which dark energy is represented by a generalized three-form field and dark matter by dust. By assuming the functions $N$ and $I$ in the model's Lagrangian as two power-law functions of the three-form field, we obtain two fixed points of the autonomous system of evolution equations, consisting of a attractor and a tracking saddle point which can be used to alleviate the coincidence problem. After marginalizing the present three-form field $\kappa X_{0}$ which is unable to be strictly restricted, we confront the model with the latest Type Ia Supernova (SN \uppercase\expandafter{\romannumeral1}a), Baryon Acoustic Oscillations (BAO) and Cosmic Microwave Backround (CMB) radiation observations with the fitting results $\Omega_{m0}= 0.280_{-0.048}^{+0.048}$ and $\lambda=0.011_{-0.032}^{+0.032}$  in the $2\sigma$ confidence level, we also find that the best fitting effective dark energy equation of state (EOS) crosses $ -1$ at redshift around 0.2.
}
\begin{document}
\maketitle

\section{Introduction}
\label{intro}

In 1998, two groups\cite{riess1998,perlmutter1999measurements}independently showed the accelerating expansion of the universe basing on Type Ia Supernova(SN \uppercase\expandafter{\romannumeral1}a) observations. Six year later, Riess et al.\cite{Riess2004Type} provided evidence at $>99\%$ for the existence of a transition for universe from deceleration to acceleration. Since all usual types of matter with positive pressure decelerate the expansion of the universe, a sector with negative pressure named as dark energy was suggested to account for the invisible fuel that accelerates the expansion rate of the current universe\cite{Sahni2004Dark,Carroll2000The}.

The $\Lambda CDM$ model is a mathematically simple and observationally consistent model, in which vacuum
energy denoted by cosmological constant $\Lambda$ plays the role of
dark energy. Although $\Lambda CDM$  model provides an excellent
fit to a wide range of astronomical data so far, such model in fact is theoretical problematic
because of two sharp puzzles, i.e., the so-called fine-tuning and coincidence problems, the
former indicates the enormous disagreement between the observational vacuum density and
the theoretical one, while the latter questions that why is the observational vacuum density
coincidentally comparable with the critical density at the present epoch in the long history of the Universe. To alleviate these two puzzles, other alternative scenarios have been
proposed, including scalar field models such as quintessence\cite{Caldwell1998Cosmological,PhysRevD.59.123504}, phantom\cite{Caldwell1999A,Carroll2003Can,Singh2003Cosmological}and k-essence\cite{Chiba1999Kinetically}.
Quintessence is considered as a dark energy scenario with EOS $>-1$ which can be described by a canonical scalar field.
Phantom is a scalar field with a negative kinetic term, it's EOS can be in a region where $\omega<-1$.
And k-essence is constructed by using kinetic term in the domain of a general function in the field Lagrangian. This model can realize both $-1$ and $1$ due to the existence of a positive energy density.

Since the experimental evidences of cosmology-specific scalars particles have not been discovered yet, there is no reason to exclude the possibility of some other high form field to be dark energy. Indeed, the three-form cosmology proposed in\cite{Koivisto2009Inflation1,Koivisto2009Three} could be a good alternative to scalar cosmology, because such high form field not only respects the Friedmann-Robertson-Walker(FRW) symmetry naturally but also can accelerates the expansion rate of the current universe without a slow-roll condition.

To alleviate the coincidence problem, our previous work \cite{Yao2018} considered a model of canonical three-form dark energy coupled to spinor dark matter. Although such model already can alleviate the coincidence problem, it's interesting to consider a coupled noncanonical three-form dark energy model. Noncanonical three-form field, also called as generalized three-form field, has already introduced in\cite{PhysRevD.96.023516,Wongjun2017Generalized} to compare with k-essence.
Also, since treating dark matter as a spinor field will bring some observational difficulty\cite{D2016Quantum}, in this paper, dark matter is represented by dust.

The contents of this paper are as follows. In section \ref{sec:1}, we present a Lagrangian describing the interaction between a generalized three-form field and point particles and then derive the field equations from such Lagrangian. In section \ref{sec:2}, we consider these field equations in a FRW
space-time by assuming the function $N$ and $M$ in the Lagrangian as two power-law functions of the three-form field and carry out a likelihood analysis of the model with the use of 1049 SN \uppercase\expandafter{\romannumeral1}a data points from recently released Pantheon \cite{Scolnic2017The} and BAO data from the WiggleZ Survey\cite{Chris2011The}, SDSS DR7 Galaxy sample \cite{Percival2010Baryon}and 6dF Galaxy Survey datasets\cite{Beutler2011The}, together with CMB data from WMAP7 observations\cite{jarosik2011seven}. In the last section, we make a brief conclusion with this paper.

\section{A field theory of generalized three-form and dust}
\label{sec:1}

A Lagrangian which describe the interaction between a noncanonical three-form field $A_{\alpha\beta\gamma}$ and dust in a curve space-time can be constructed as
\begin{equation}\label{}
  \mathcal{L}_{m}=-\frac{1}{48}F^{2}N(A^{2})-I(A^{2})\tilde{\rho}_{m}+\lambda_{1}(g_{\mu\nu}u^{\mu}u^{\nu}+1)+\lambda_{2}\nabla_{\alpha}(\tilde{\rho}_{m}u^{\alpha})
\end{equation}
where $F=dA$ denotes the field strength tensor, $N(A^{2}) $ represents the function that noncanonicalize the three-form field, $I(A^{2})$ represents the coupling function, and $\tilde{\rho}_{m}$ is the density of the decoupled dark matter, $u^{\alpha}$ is the 4-velocity of dark matter, $\lambda_{1}$ and $\lambda_{2}$ are multipliers.

 We can obtains the dynamics field equations from the total action
\begin{equation}\label{}
 S=\int\mathcal{L}\sqrt{-g}d^{4}x
\end{equation}
where $\mathcal{L}=\mathcal{L}_{g}+\mathcal{L}_{m}=\frac{R}{2\kappa^{2}}+\mathcal{L}_{m}$ is the Lagrangian including gravity and matter,
$R$ denotes the Ricci scalar and $\kappa=\sqrt{8\pi G}$ is the inverse of the reduced Planck mass.

The variation of the action with respect to the $g^{\mu\nu},\tilde{\rho}_{m},u^{\alpha},\lambda_{1},\lambda_{2}$ leads to
\begin{eqnarray}
R_{\mu\nu}-\frac{1}{2}g_{\mu\nu}R &=& \kappa^{2}T_{\mu\nu} \\
(\partial_{\alpha}\lambda_{2})u^{\alpha} &=& -I \\
\lambda_{1} &=& \frac{1}{2}I\tilde{\rho}_{m} \\
g_{\mu\nu}u^{\mu}u^{\nu}+1 &=& 0 \\
\nabla_{\alpha}(\tilde{\rho}_{m}u^{\alpha}) &=& 0
\end{eqnarray}
with the help of equation (4) and (5), the total energy-momentum tensor for three-form field and dust is written as
\begin{equation}\label{}
\begin{split}
 T_{\mu\nu}=&\frac{1}{6}NF_{\mu\alpha\beta\gamma}F_{\nu}^{\alpha\beta\gamma}+6(\frac{1}{48}\frac{dN}{dA^{2}}F^{2}+\frac{dlnI}{dA^{2}}\rho_{m})A_{\mu}^{\alpha\beta}A_{\nu\alpha\beta}\\
 &-g_{\mu\nu}\frac{1}{48}F^{2}N+\rho_{m}u_{\mu}u_{\nu}.
\end{split}
\end{equation}
We have defined the dark matter density to be
\begin{equation}
  \rho_{m}=I(A^{2})\tilde{\rho}_{m}
\end{equation}
By varying the total action with respect to the three-form field, we have the following equations of motion
\begin{equation}\label{}
  (\nabla_{\alpha}N)F^{\alpha\mu\nu\rho}+N\nabla_{\alpha}F^{\alpha\mu\nu\rho} = 12(\frac{1}{48}\frac{dN}{dA^{2}}F^{2}+\frac{dlnI}{dA^{2}}\rho_{m})A^{\mu\nu\rho}
\end{equation}
Noting that $N$ is a function of $A^2$ instead of $F^2$, so all of the second derivatives terms contain in the term  $N\nabla_{\alpha}F^{\alpha\mu\nu\rho}$, therefore the noncanonical field theory proposed in the paper is satisfied with Hyperbolicity condition in the same way with the canonical field theory.

Using the equations of motion for the three-form field and the vanishing of the divergence of the total stress energy tensor we have the
equation of motion for the dust:
\begin{equation}\label{}
  \nabla_{\mu}(\rho_{m}u^{\mu}u_{\nu})=-2\frac{dlnI}{dA^{2}}\rho_{m}A^{\alpha\beta\gamma}\nabla_{\nu}A_{\alpha\beta\gamma}
\end{equation}

\section{A coupled generalized three-form dark energy model}
\label{sec:2}
\subsection{Cosmological evolution of a coupled generalized three-form dark energy model}
Now we can consider the field equations in a homogeneous, isotropic, and spatially flat space-time, it is described by the following metric
\begin{equation}\label{}
  ds^{2}=-dt^{2}+a(t)^{2}d\vec{x}^{2}
\end{equation}
here $a(t)$ stands for the scale factor.

The three-form field is assumed as the following time-like component of the dual vector field for the purpose to be compatible with FRW symmetries
\begin{equation}
  A_{i j k}=X(t)a(t)^{3}\varepsilon_{ijk}
\end{equation}
To construct a coupled generalized three-form dark energy model, we choose the function $N$ and function $I$ to be
\begin{eqnarray}\label{}
  N &=&[(\frac{\kappa^{2}}{6}A^{2})]^{\frac{\lambda}{2}}=(\kappa \mid X \mid)^{\lambda}   \\
  I &=& N
\end{eqnarray}
\footnote{ For simplicity, we consider $X > 0$ and neglect the absolute value sign in the following discussion.}
where $\lambda$ are constants.

After then, we have the Friedmann equations
\begin{eqnarray}
  H^{2} &=& \frac{\kappa^{2}}{3}\rho \\
  \dot{H} &=&-\frac{\kappa^{2}}{2}(\rho+p)
\end{eqnarray}
with
\begin{eqnarray}
  \rho &=&-T_{0}^{0}=-g^{00}T_{00}=\frac{1}{2}(\kappa X)^{\lambda}(\dot{X}+3HX)^{2}+\rho_{m} \\
  p &=&\frac{1}{3}T_{i}^{i}=\frac{1}{3}g^{ii}T_{ii}=-\frac{1+\lambda}{2}(\kappa X)^{\lambda}(\dot{X}+3HX)^{2}+\lambda \rho_{m}
\end{eqnarray}
In the FRW space-time, there is only one independent equation of motion of the three-form field
\begin{equation}\label{}
  (\kappa X)^{\lambda}(\ddot{X}+3\dot{H}X+3H\dot{X})+\frac{\lambda \kappa (\kappa X)^{\lambda-1}}{2}(\dot{X}-3HX)(\dot{X}+3HX)+ \frac{\lambda\rho_{m}}{X}=0
\end{equation}

and the continuity equations of two dark sector is
\begin{eqnarray}
  \dot{\rho}_{X}+3H(\rho_{X}+p_{X}) &=& -\delta H\rho_{m} \\
  \dot{\rho}_{m}+3H(\rho_{m}+p_{m}) &=& \delta H\rho_{m}
\end{eqnarray}
with
\begin{eqnarray}
  \rho_{X} &=& \frac{1}{2}(\kappa X)^{\lambda}(3HX+\dot{X})^{2}, \hspace{1cm}   p_{X}=p
\end{eqnarray}
\begin{equation}\label{}
\delta=\lambda\frac{X^{\prime}}{X}
\end{equation}
the prime stands for derivative with respect to e-folding time $N=\ln a$ here and in the following.

\begin{figure}[h]
%\begin{tabular}{cc}
\begin{minipage}{0.4\linewidth}
  \centerline{\includegraphics[width=1\textwidth]{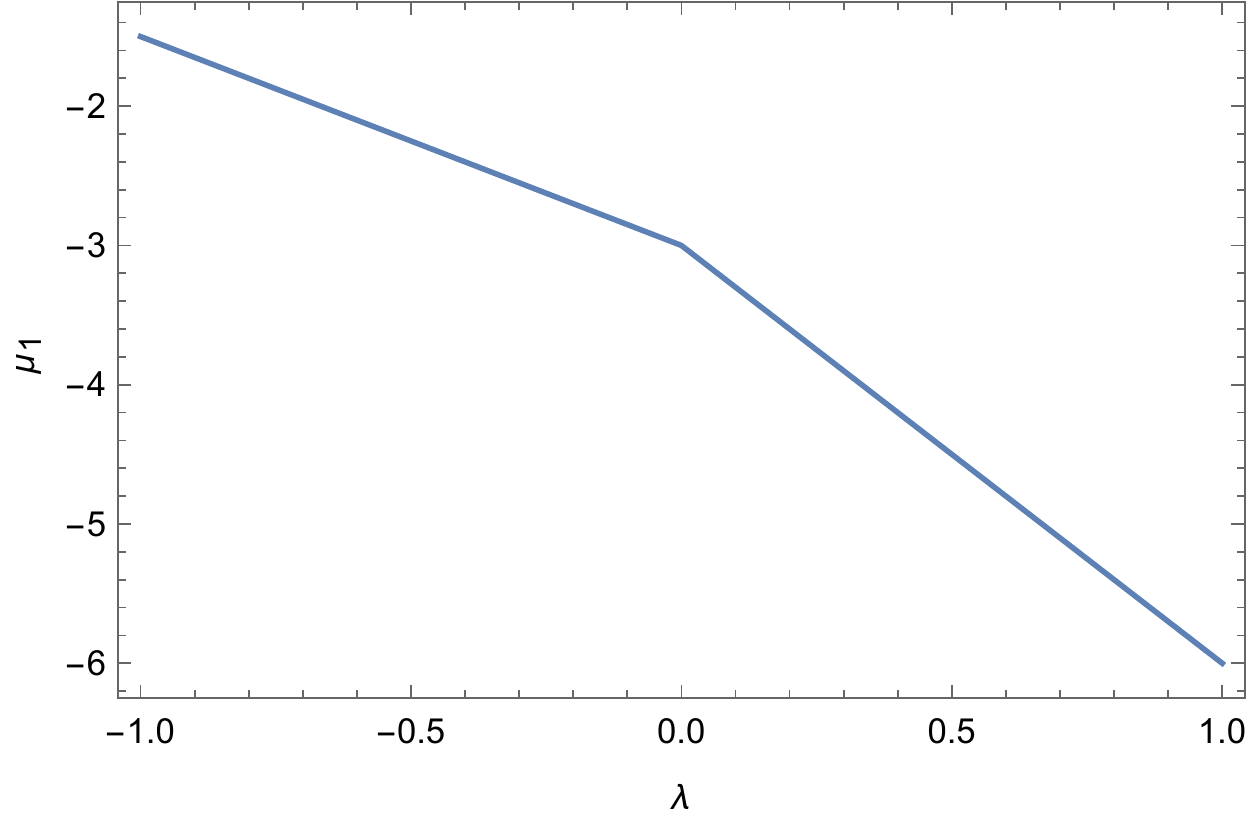}}
  \label{a}
\end{minipage}
\begin{minipage}{0.4\linewidth}
  \centerline{\includegraphics[width=1\textwidth]{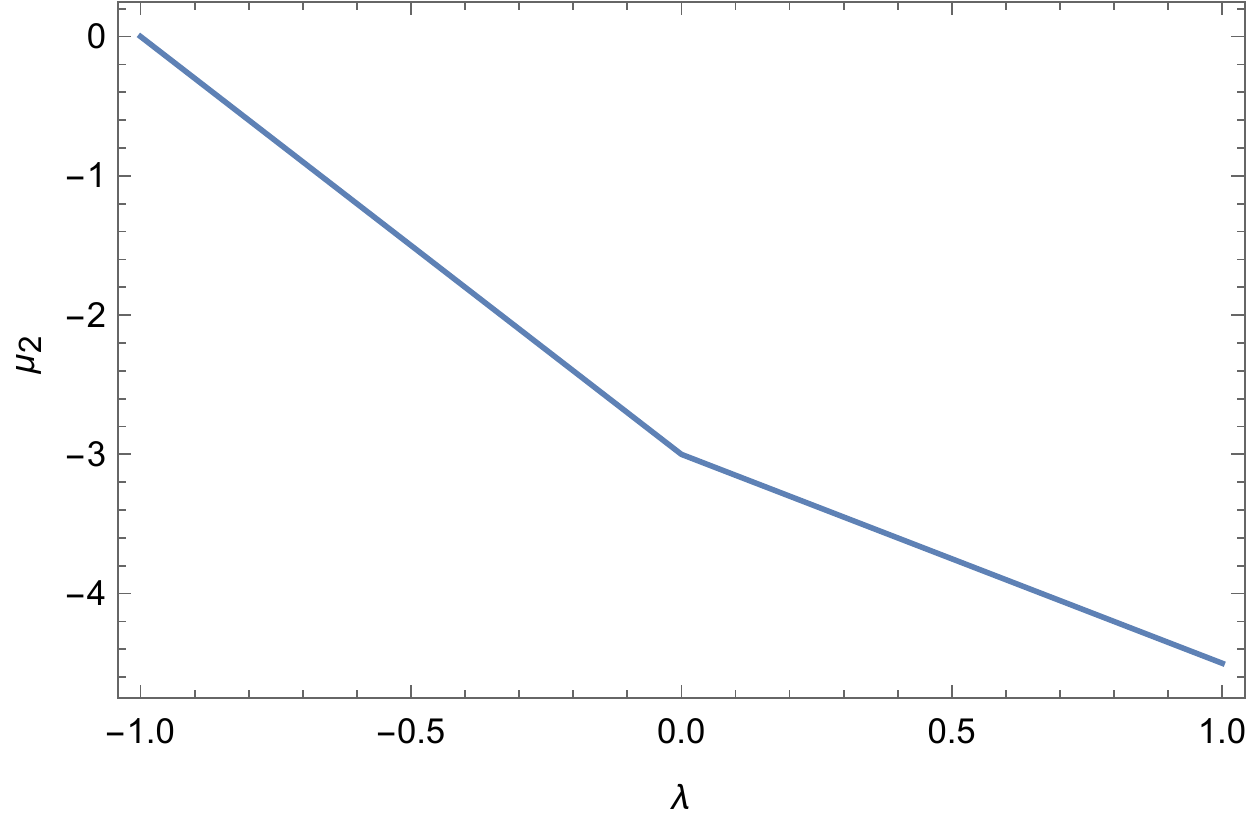}}
\end{minipage}
\caption{Two eigenvalues with respect to $\lambda$ of the first fixed point.}
\label{fig:1}
\end{figure}

In order to study dynamics behaviors of the universe, it is convenient to introduce the following dimensionless variable
\begin{equation}\label{}
  x=\kappa X, \hspace{1cm} y=\frac{\kappa}{\sqrt{6}}(X^{\prime}+3X)
\end{equation}
By applying the Friedmann equations and equations of motion, one can obtains the autonomous system of evolution equations
\begin{eqnarray}
  x^{\prime} &=& \sqrt{6}y-3x\\
  y^{\prime} &=& \frac{3}{2}x^{\lambda}(-\lambda y^{2}+(1+\lambda)(x^{-\lambda}- y^{2}))y+3\lambda y-\frac{\sqrt{6}}{2} \frac{\lambda}{x}x^{-\lambda}.
\end{eqnarray}
There are two fixed points for such autonomous system. One of them is $\left((\frac{2}{3})^{\frac{1}{2+\lambda}},(\frac{3}{2})^{\frac{1}{2}-\frac{1}{2+\lambda}}\right)$, As is shown in the Fig.\ref{fig:1}, it's eigenvalues $(\mu_{1}(\lambda),\mu_{2}(\lambda))$ are both negative when $-1<\lambda<1$, so in this interval, such fixed point is an attractor.
 By rewriting density and pressure in term of the dimensionless variables, we have
\begin{equation}\label{}
  \Omega_{X}=x^{\lambda}y^{2}=1
\end{equation}
\begin{equation}\label{}
  \omega_{X}=\frac{p_{X}}{\rho_{X}}=-1+\lambda \frac{x^{-\lambda}-2y^{2}}{y^{2}}=-1-\lambda
\end{equation}
indicating that such fixed point represents a three-form saturated universe .

\begin{figure}[h]
%\begin{tabular}{cc}
\begin{minipage}{0.4\linewidth}
  \centerline{\includegraphics[width=1\textwidth]{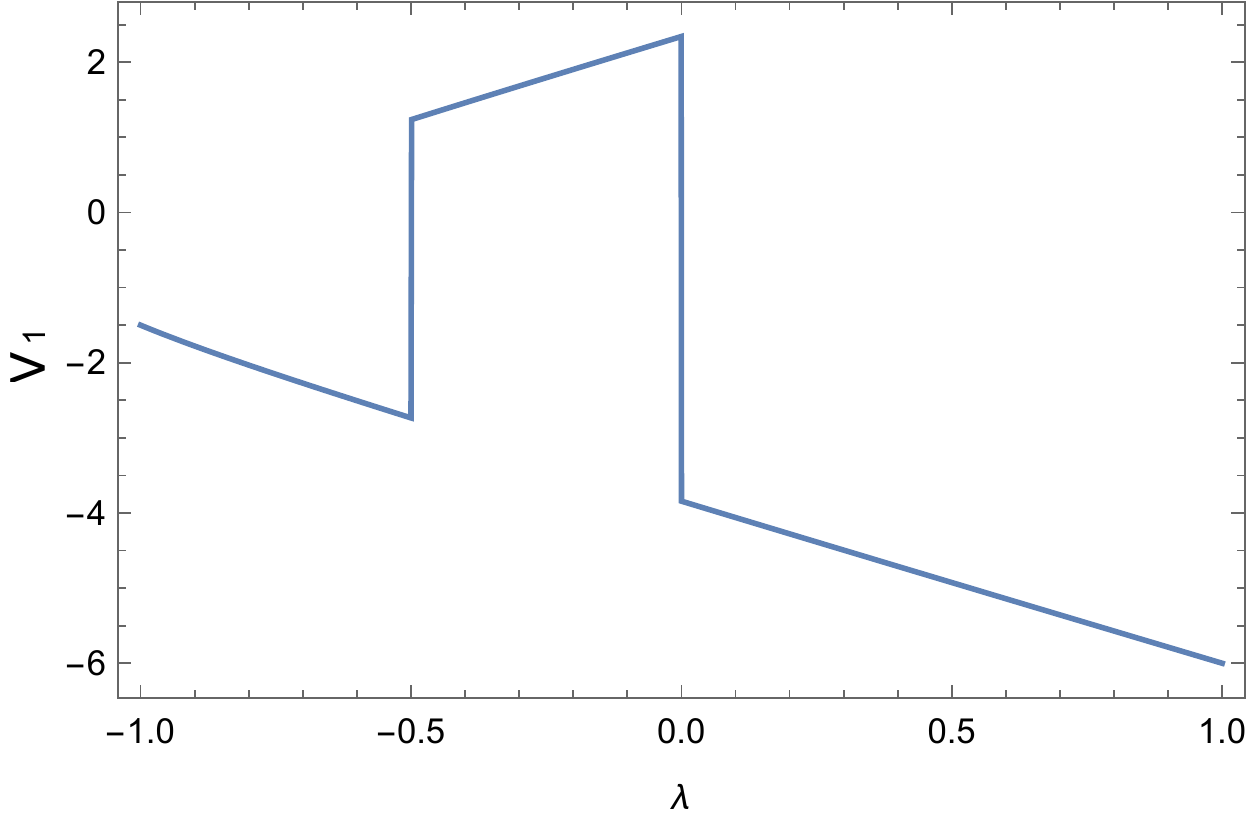}}
  \label{a}
\end{minipage}
\begin{minipage}{0.4\linewidth}
  \centerline{\includegraphics[width=1\textwidth]{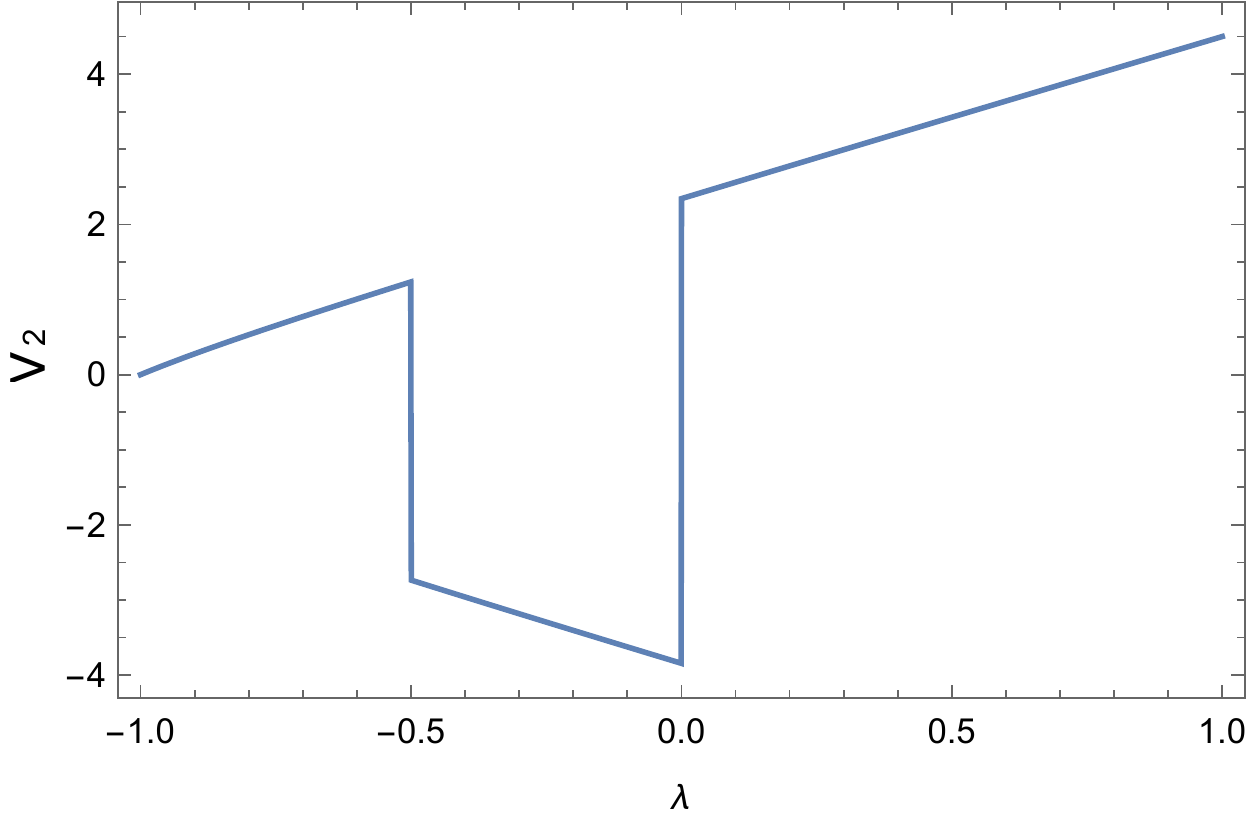}}
\end{minipage}
\caption{Two eigenvalues with respect to $\lambda$ of the second fixed point.}
\label{fig:2}
\end{figure}

The other one is $\left(2^{\frac{1}{2+\lambda}}(\frac{3+6\lambda}{\lambda})^{-\frac{1}{2+\lambda}},2^{-\frac{1}{2}+\frac{1}{2+\lambda}}\sqrt{3}(\frac{3+6\lambda}{\lambda})^{-\frac{1}{2+\lambda}}\right)$, which is a saddle point as long as $-1<\lambda<1$, it's eigenvalues $\left(\nu_{1}(\lambda),\nu_{2}(\lambda)\right)$ are shown in Fig.\ref{fig:2}.
It can be inferred from
\begin{equation}\label{}
  \omega_{X}=\frac{p_{X}}{\rho_{X}}=-1+\lambda \frac{x^{-\lambda}-2y^{2}}{y^{2}}=0   , \delta=\lambda \frac{x'}{x}=0
\end{equation}
  that such fixed point is a tracking solution which can be used to alleviate the coincidence problem with a fine-turning of the model parameters.

With the help of the equations (14),(15), we can derive $(x^{\lambda}y^{2})^{\prime}\mid_{x=0^{+},y=0}=-3\lambda\neq3\lambda=(x^{\lambda}y^{2})^{\prime}\mid_{x=0^{-},y=0}(\lambda\neq0)$ from the dynamical equations, which suggests that this model is just a simplified model since $x=0,y=0$ is a non-differentiable point for $x^{\lambda}y^{2}$. We expect that in a real model, $N$ and $I$ are replaced by a differentiable piecewise function that is equal to $N$ and $I$  when $x\in(-\infty,-\varepsilon)\cup(\varepsilon,\infty)$ ($\varepsilon$ is  a tiny constant), which leading to $(x^{\lambda}y^{2})^{\prime}\mid_{x=0^{+},y=0}=0=(x^{\lambda}y^{2})^{\prime}\mid_{x=0^{-},y=0}$. Therefore, in such a real model, $x=0,y=0$ is another fixed point. However, as long as $x\in(-\infty,-\varepsilon)\cup(\varepsilon,\infty)$ for the real cosmological solution, our simplified model is reliable.

\begin{figure}
  \includegraphics[width=0.55\textwidth]{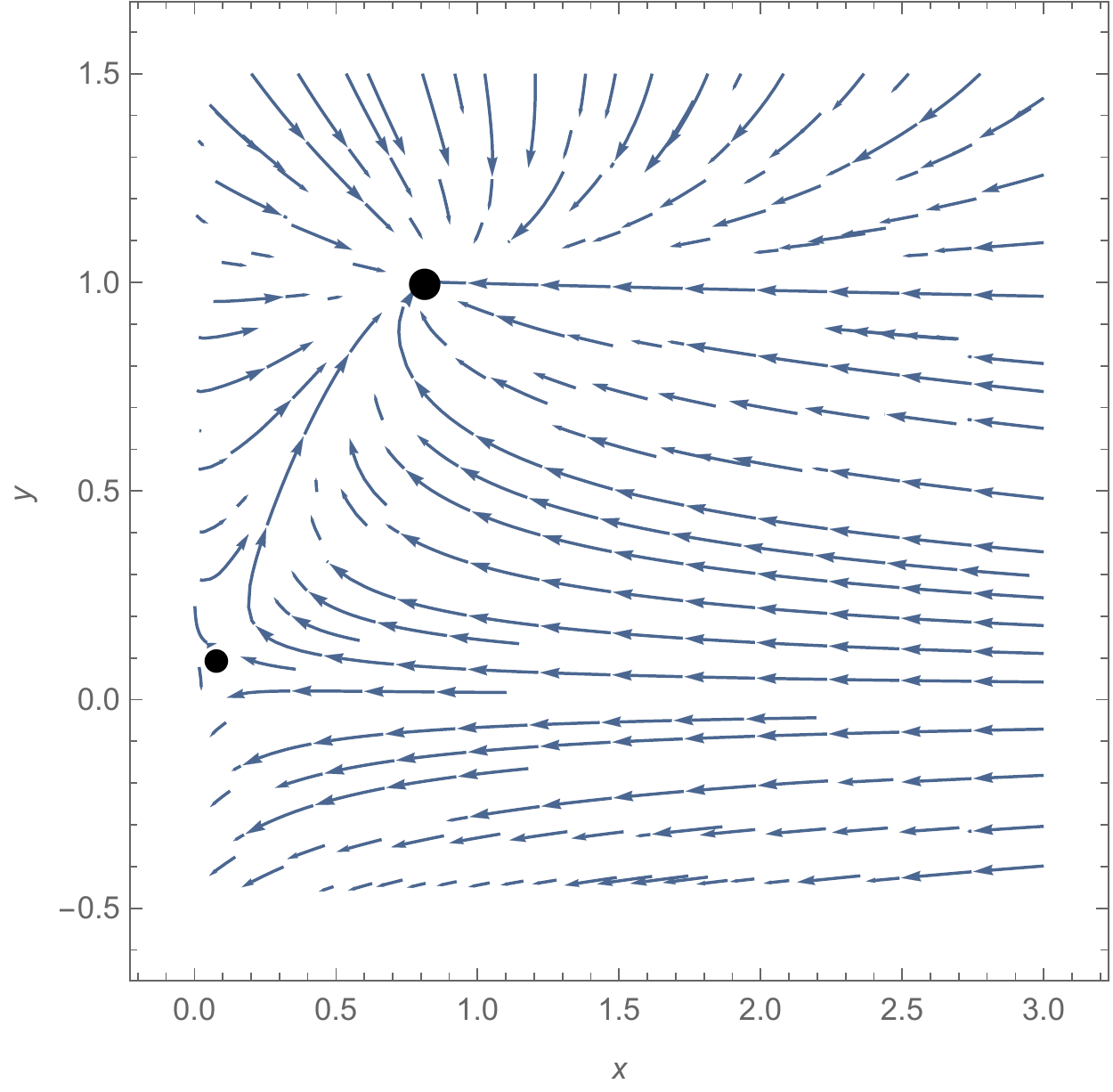}
\caption{The larger black point represents the attractor, and the smaller one represents the saddle point, blue lines with arrow represent phase space trajectories.}
\label{fig:3}       % Give a unique label
\end{figure}

The trajectories with respect to $x(N)$ and $y(N)$ with a wide range of initial conditions and a assumption that $\lambda=0.01$ (we will see that this is a good choice in the next section) can be visualized by Fig.\ref{fig:3}, which shows that trajectories run toward attractor, coasting along the saddle point.

\subsection{Confront the coupled generalized three-form dark energy model with observations}
\label{sec:3}
In this section, we perform a likelihood analysis on the model parameters with the combination of data from SN \uppercase\expandafter{\romannumeral1}a, BAO and CMB radiation observations.

First of all, we construct the $ \chi^{2}$ function for SN \uppercase\expandafter{\romannumeral1}a by using the 1049 Pantheon data points as
\begin{equation}\label{}
  \chi_{SN \uppercase\expandafter{\romannumeral1}a}^{2}= P- \frac{Q^{2}}{R}
\end{equation}
where $P$, $Q$ and $R$ are defined as
\begin{eqnarray}
  P &=& \sum_{i=1}^{1049}\frac{\left(\mu_{th}(z_{i})-\mu_{obs}(z_{i})\right)^{2}}{\sigma_{\mu}^{2}(z_{i})},\\
  Q &=& \sum_{i=1}^{1049}\frac{\left(\mu_{th}(z_{i})-\mu_{obs}(z_{i})\right)}{\sigma_{\mu}^{2}(z_{i})}, \\
  R &=& \sum_{i=1}^{1049}\frac{1}{\sigma_{\mu}^{2}(z_{i})},
\end{eqnarray}
here $\mu_{th}=5\log_{10}\left[(1+z)\int_{0}^{z}\frac{H_{0}}{H(z^{\prime})}dz^{\prime}\right]+25$ denotes the theoretical distance modulus  and $\mu_{obs}$ stands for the observed one and have a statistical uncertainty  $\sigma_{\mu}$.

In the second step, we consider BAO data from the WiggleZ Survey, SDSS DR7 Galaxy sample and 6dF Galaxy Survey together with CMB data from WMAP 7 yeas observations to obtain the BAO/CMB constraints on the model parameters by defining $\chi_{BAO/CMB}^{2}$ as\cite{Giostri2012From,Mamon2016Constraints,Mamon2016A}

\begin{equation}\label{}
  \chi_{BAO/CMB}^{2}=X^{T}C^{-1}X
\end{equation}
where
\begin{equation}\label{}
X=\left(
\begin{matrix}
\frac{d_{A}(z_{*})}{D_{v}(0.106)}-30.95\\
\frac{d_{A}(z_{*})}{D_{v}(0.2)}-17.55\\
\frac{d_{A}(z_{*})}{D_{v}(0.35)}-10.11\\
\frac{d_{A}(z_{*})}{D_{v}(0.44)}-8.44\\
\frac{d_{A}(z_{*})}{D_{v}(0.6)}-6.69\\
\frac{d_{A}(z_{*})}{D_{v}(0.73)}-5.45
\end{matrix}
\right)
\end{equation}
in which $d_{A}(z)=\int_{0}^{z}\frac{1}{H(z^{\prime})}dz^{\prime}$ and $D_{V}(z)=\left[d_{A}(z)^{2}\frac{z}{H(z)}\right]^{\frac{1}{3}}$ represent the co-moving angular-diameter distance and the dilation scale respectively, while $z_{*}\approx1091$ denotes the decoupling time.
\begin{equation*}
C^{-1}=\left(
\begin{matrix}
  0.48435 & -0.101383 & -0.164945 & -0.0305703 & -0.097874 & -0.106738 \\
  -0.101383 & 3.2882 & -2.45497 & -0.0787898 & -0.252254 & -0.2751 \\
  -0.164945 & -2.45497 & 9.55916 & -0.128187 & -0.410404 & -0.447574 \\
  -0.0305703 & -0.0787898 & -0.128187 & 2.78728 & -2.75632 & 1.16437 \\
  -0.097874 & -0.252254 & -0.410404 & -2.75632 & 14.9245 & -7.32441 \\
  -0.106738 & -0.2751 & -0.447574 & 1.16437 & -7.32441 & 14.5022
\end{matrix}
\right)
\end{equation*}
is the inverse of the correlation matrix.

Finally, the total $\chi^{2}$ function for the combined observational datasets is given by $\chi^{2}=\chi_{SN \uppercase\expandafter{\romannumeral1}a}^{2}+\chi_{BAO/CMA}^{2}$, from which we can construct the likelihood function as $L=L_{0} e^{-\frac{1}{2}\chi^{2}}$, here $L_{0}$ is a normalized constant which is independent of the parameters.

\begin{figure}
  \includegraphics[width=0.5\textwidth]{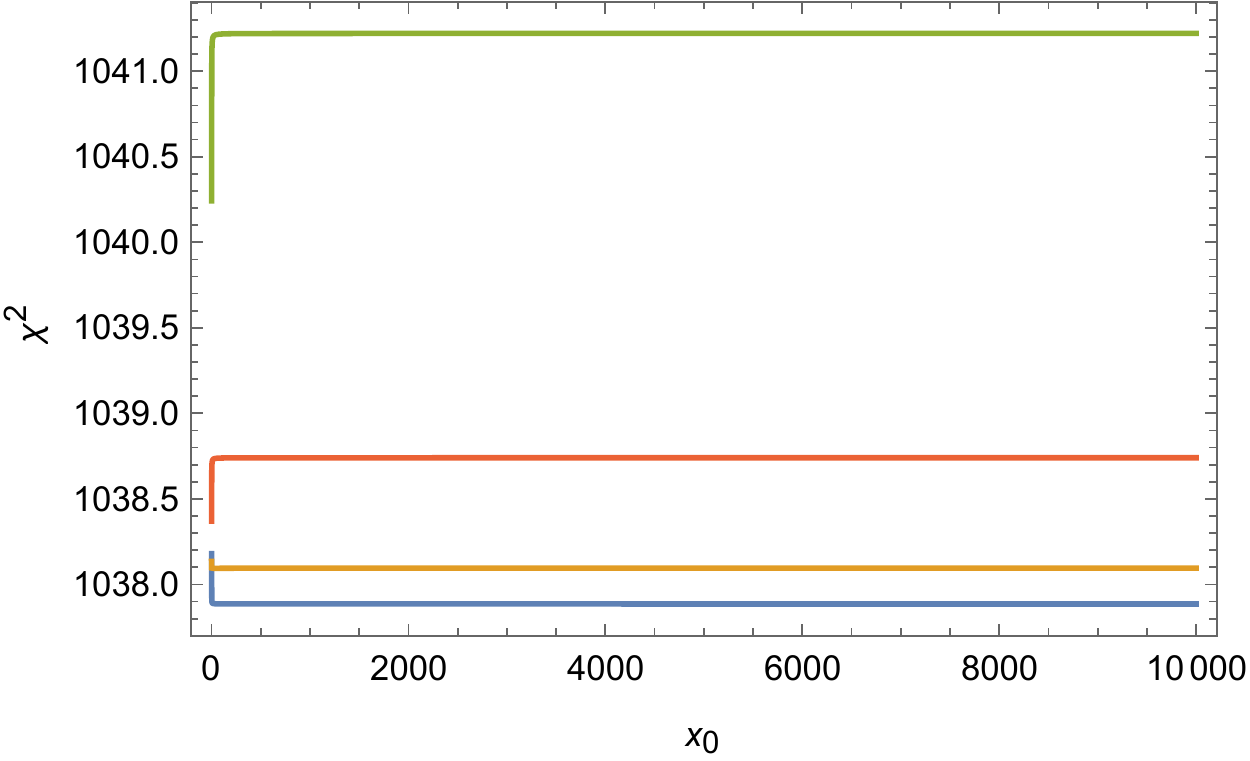}
\caption{$\chi^{2}$ with respect to $\lambda$, the green, red, orange, blue line are corresponding to the case that $(\lambda,\Omega_{m0})=(0.011,0.29),(0.011,0.3),(0.015,0.28),(0.011,0.28)$.}
\label{fig:4}       % Give a unique label
\end{figure}

\begin{figure}[h]
%\begin{tabular}{cc}
\begin{minipage}{0.3\linewidth}
  \centerline{\includegraphics[width=1\textwidth]{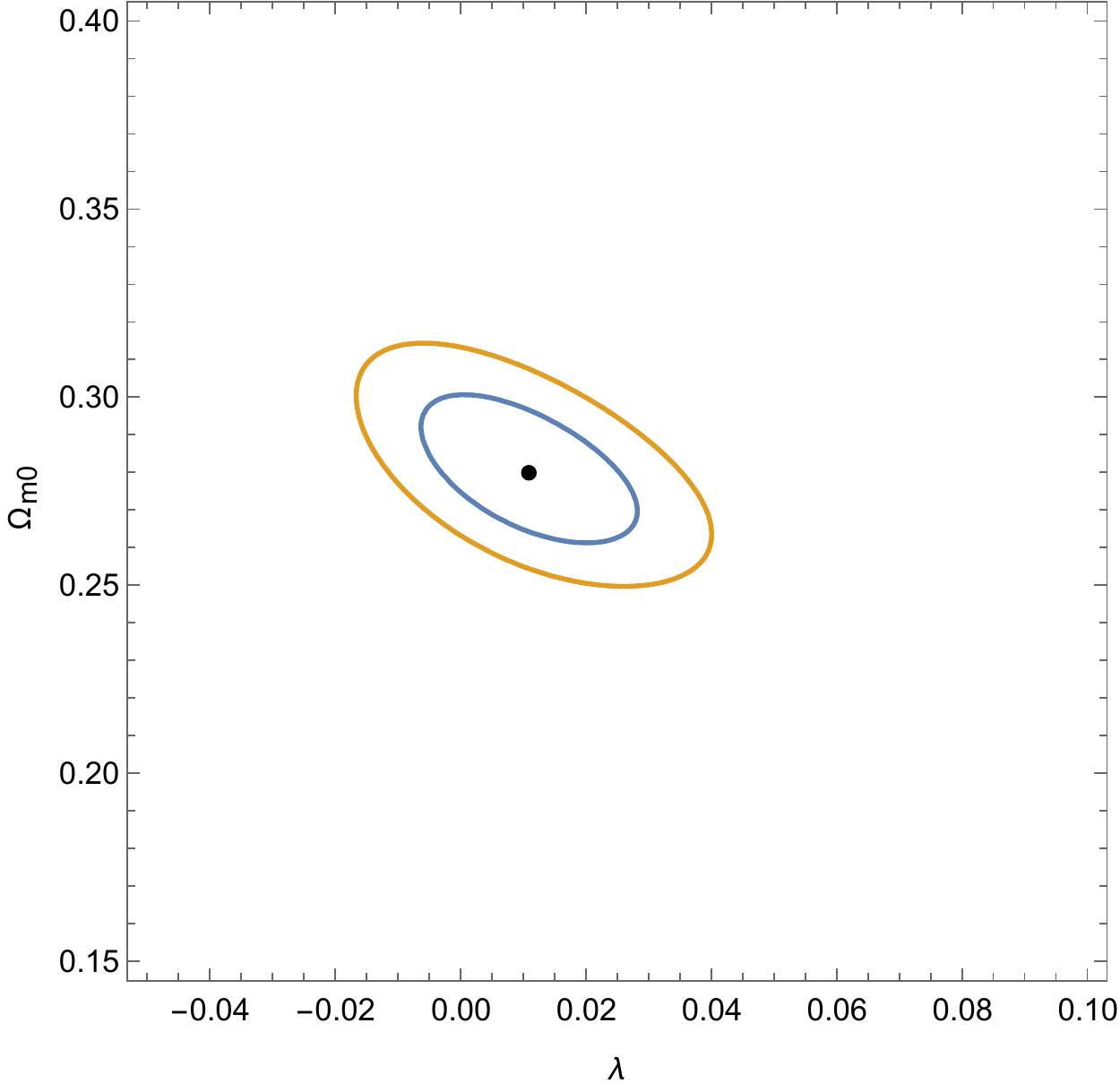}}
  \label{a}
\end{minipage}
\begin{minipage}{0.3\linewidth}
  \centerline{\includegraphics[width=1\textwidth]{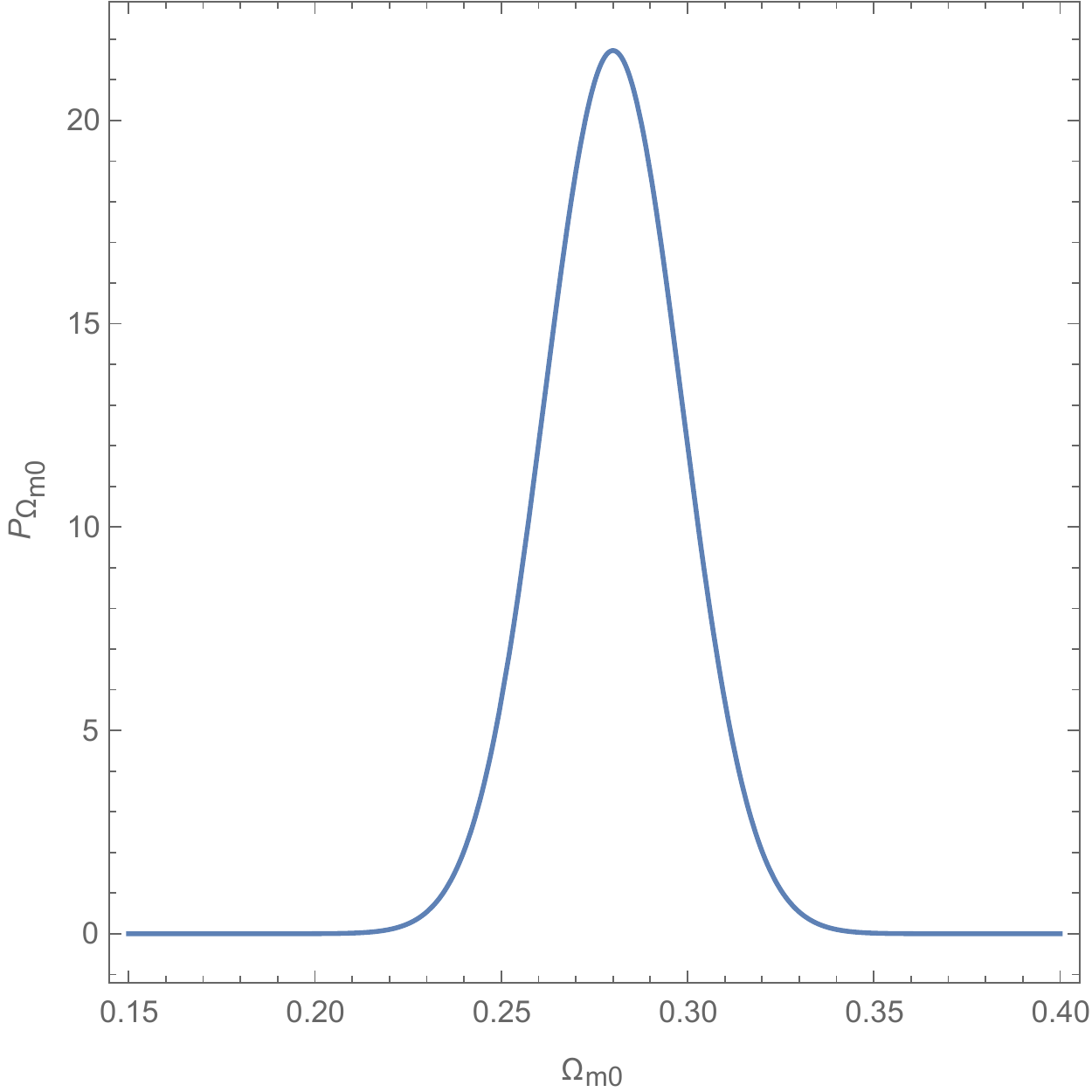}}
  \label{a}
\end{minipage}
\begin{minipage}{0.3\linewidth}
  \centerline{\includegraphics[width=1\textwidth]{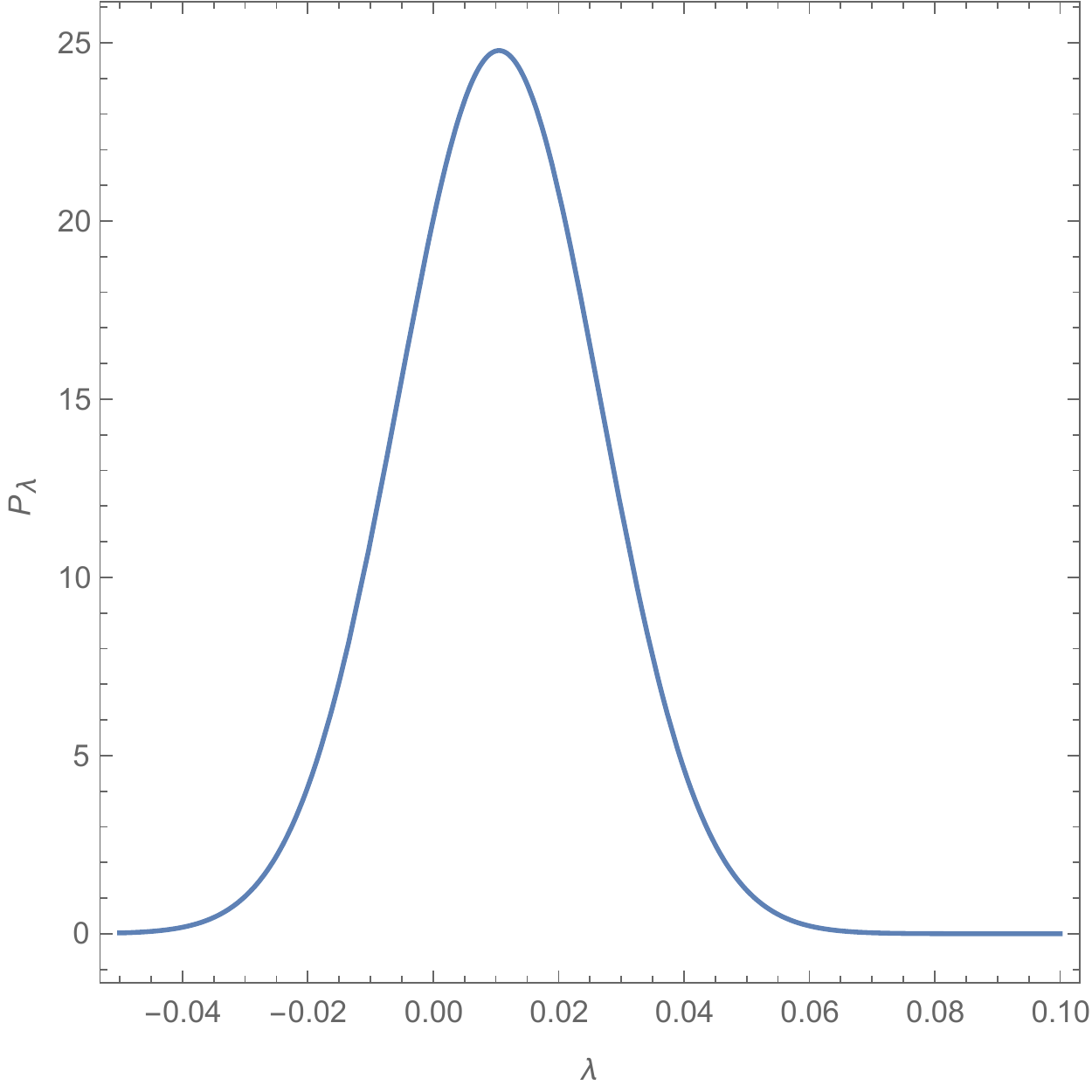}}
\end{minipage}
\caption{ The first figure shows observational constraints on parameters $(\lambda,\Omega_{m0})$ with the
combination of SN Ia and BAO/CMB detasets, in which the region inside orange line and blue line are corresponding to $2\sigma$ and $1\sigma$ region respectively, while the black point (0.011,0.28)
represents the best-fit value of the pair $(\lambda,\Omega_{m0})$. The second figure is the likelihood function
of $\Omega_{m0}$ which is marginalized with a flat prior that becomes zero if $\lambda$ bigger than 0.1 or smaller
than $-0.05$ .The third figure is the likelihood function
of $\lambda$ which is marginalized with a flat prior that becomes zero if $\Omega_{m0}$ bigger than 0.4 or smaller
than 0.15.}
\label{fig:5}
\end{figure}

Providing with the likelihood function, one can then obtain the best-fit values of the free parameters by maximizing it. However, it can be inferred from Fig.\ref{fig:4} that the parameter $x_{0}$ (the present value of $x$) can't be strictly restricted, the figure also shows that the likelihood function becomes a none zero constant with respect to $x_{0}$ when $x_{0}$ is large enough.
Given this behavior of the likelihood function $L$, we have such formula
\begin{equation}\label{}
  \int_{0}^{+\infty}L(x_{0},\lambda,\Omega_{m0})dx_{0} \propto \lim_{x_{0}\rightarrow+\infty}L(x_{0},\lambda,\Omega_{m0})
\end{equation}
therefore, the likelihood function that has been marginalized over $x_{0}$ without a prior can be expressed as
 \begin{equation}\label{}
 \lim_{x_{0}\rightarrow+\infty}L(x_{0},\lambda,\Omega_{m0})\approx L(\tilde{x_{0}},\lambda,\Omega_{m0})
\end{equation}
 where $\tilde{x_{0}}$ is a large number, one can choose it as 10000, for example. We now present the fitting result in Fig.\ref{fig:5} and Tab.\ref{tab:1} by analyzing such likelihood.

\begin{figure}
  \includegraphics[width=0.5\textwidth]{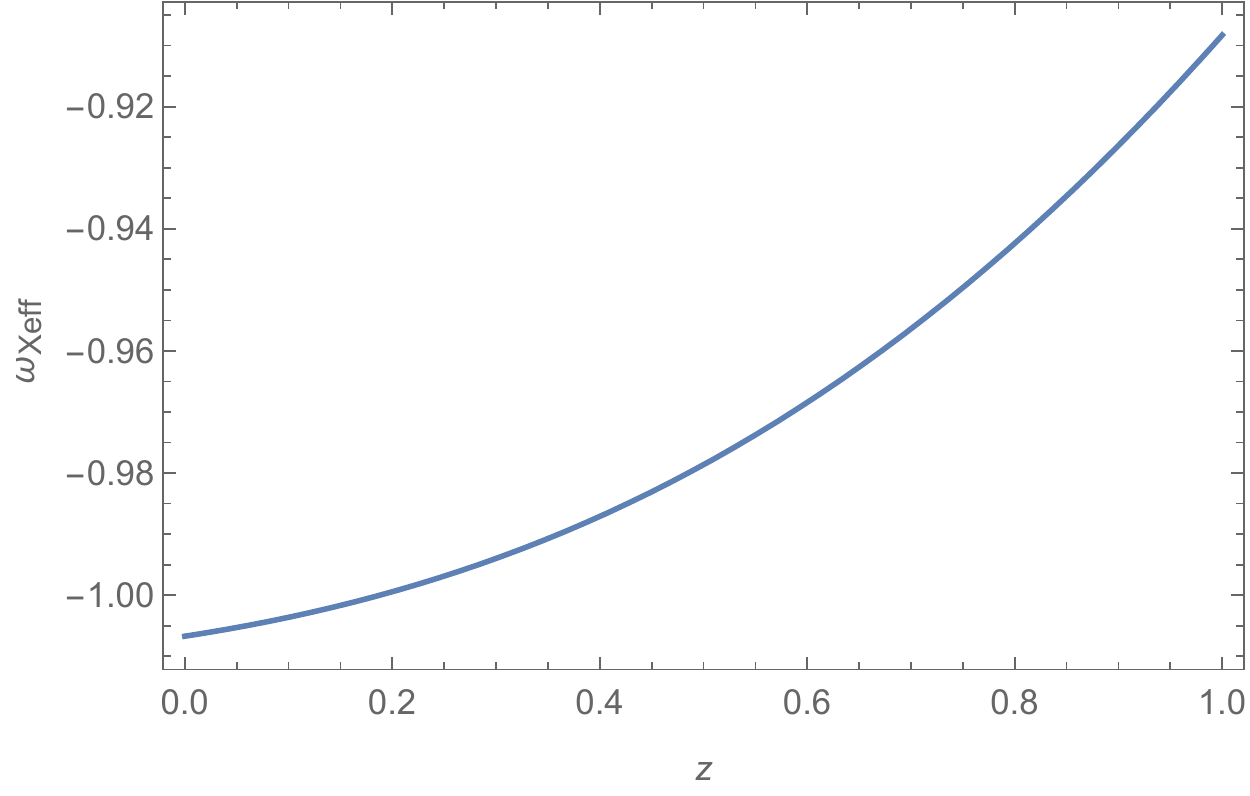}
\caption{The effective EOS of dark energy which crosses the cosmological constant boundary $-1$ at around $z=0.2$ }
\label{fig:6}       % Give a unique label
\end{figure}

 The Tab.\ref{tab:1} shows that observations favor a small positive coupling constant. Comparing with (28) and (29), we find that the predicted universe favor a phantom-like future. Confronting the $\Lambda CDM$ model with the same data set, we have $\Omega_{m0}=0.287_{-0.022}^{+0.022}$  and the $AIC=1040.76$, which suggests the standard model still slightly favor over the new model. To prove that such model leads to a crossing of the cosmological constant boundary $-1$ at low redshift for the effective EOS of dark energy, let us rewrite Hubble parameter as
\begin{equation}\label{}
\begin{split}
   \frac{H^{2}(z)}{H_{0}^{2}} &=\frac{\Omega_{m0}\left(\frac{x}{x_{0}}\right)^{\lambda}(1+z)^{3}}{1-x^{\lambda}y^{2}}  \\
     &=\Omega_{m0}(1+z)^{3}+(1-\Omega_{m0})\frac{\rho_{Xeff}}{\rho_{Xeff0}}
\end{split}
\end{equation}
where
\begin{equation}\label{}
\begin{split}
\frac{\rho_{Xeff}}{\rho_{Xeff0}} &=\frac{\Omega_{m0}}{1-\Omega_{m0}}\left(\frac{\left(\frac{x}{x_{0}}\right)^{\lambda}(1+z)^{3}}{1-x^{\lambda}y^{2}}-(1+z)^{3}\right)\\
&=\exp\left[\int_{0}^{z}\frac{3(1+\omega_{Xeff}(\tilde{z}))}{1+\tilde{z}}d\tilde{z}\right]
\end{split}
\end{equation}

is the normalized effective dark energy density and
\begin{equation}\label{}
  \omega_{Xeff}(z)=-1+\frac{1}{3}\frac{x^{\lambda}\left(3+\frac{\lambda(1+z)}{x}\frac{dx}{dz}\right)(1-x^{\lambda}y^{2})+(1+z)(y^{2}\frac{\lambda}{x}\frac{dx}{dz}+\frac{dy^{2}}{dz})x^{2\lambda}-3(1-x^{\lambda}y^{2})^{2}x_{0}^{\lambda}}{\left(x^{\lambda}-(1-x^{\lambda}y^{2})x_{0}^{\lambda}\right)(1-x^{\lambda}y^{2})}.
\end{equation}
is the effective EOS of dark energy.
Plot the best fitting effective EOS of dark energy in the Fig.\ref{fig:6}, we can see that the effective EOS of dark energy cross the cosmological constant boundary $-1$ at around $z=0.2$. In another words, the coupled generalized three-form dark energy model is a quintom model.

\begin{table}
\begin{center}
\begin{tabular}{cc|  cc }
\hline\hline parameters & &  confidence interval
\\ \hline
$\Omega_{m0}$    && $ 0.280_{-0.048}^{+0.048}$
                     \\
$\lambda$         &&  $0.011_{-0.032}^{+0.032}$
                     \\
 \hline $AIC$  &&  $1041.88$
                      \\
\hline\hline
\end{tabular}
\caption{$2\sigma$ confidence interval of the model parameters and the AIC of the model.}
\label{tab:1}
\end{center}
\end{table}
\subsection{Coupled canonical three-form dark energy model as comparison}
\begin{figure}[h]
%\begin{tabular}{cc}
\begin{minipage}{0.3\linewidth}
  \centerline{\includegraphics[width=1\textwidth]{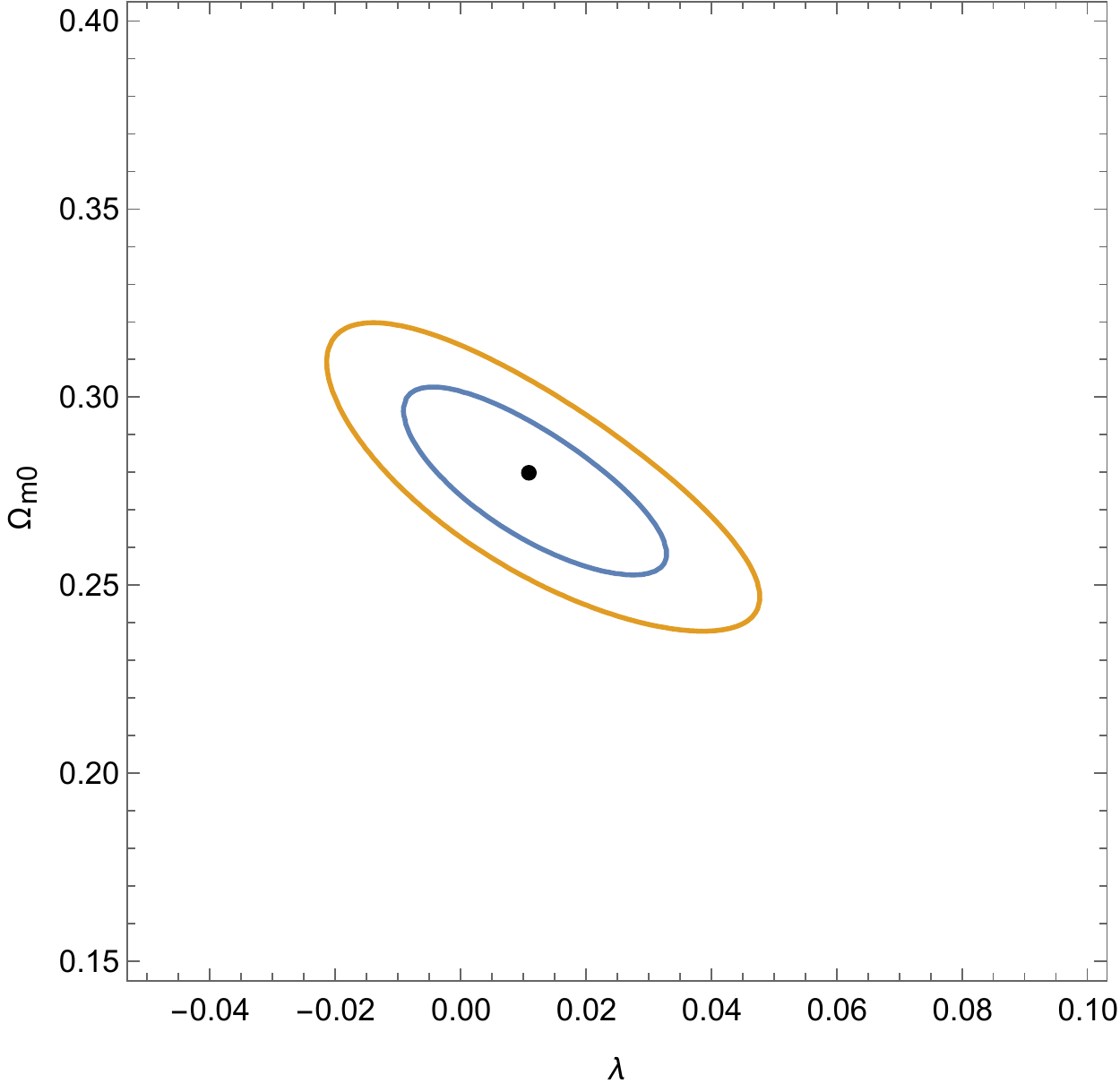}}
  \label{a}
\end{minipage}
\begin{minipage}{0.3\linewidth}
  \centerline{\includegraphics[width=1\textwidth]{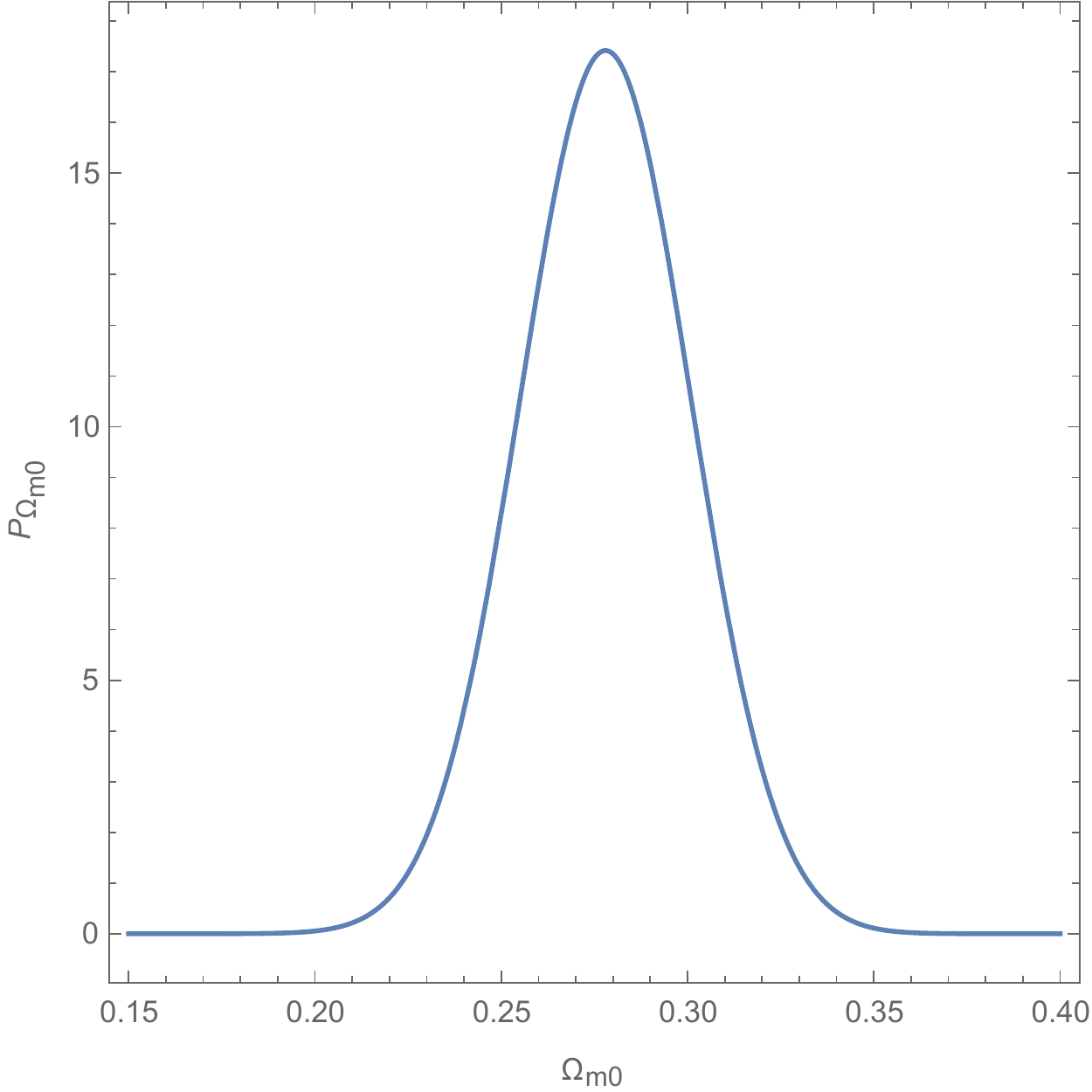}}
  \label{a}
\end{minipage}
\begin{minipage}{0.3\linewidth}
  \centerline{\includegraphics[width=1\textwidth]{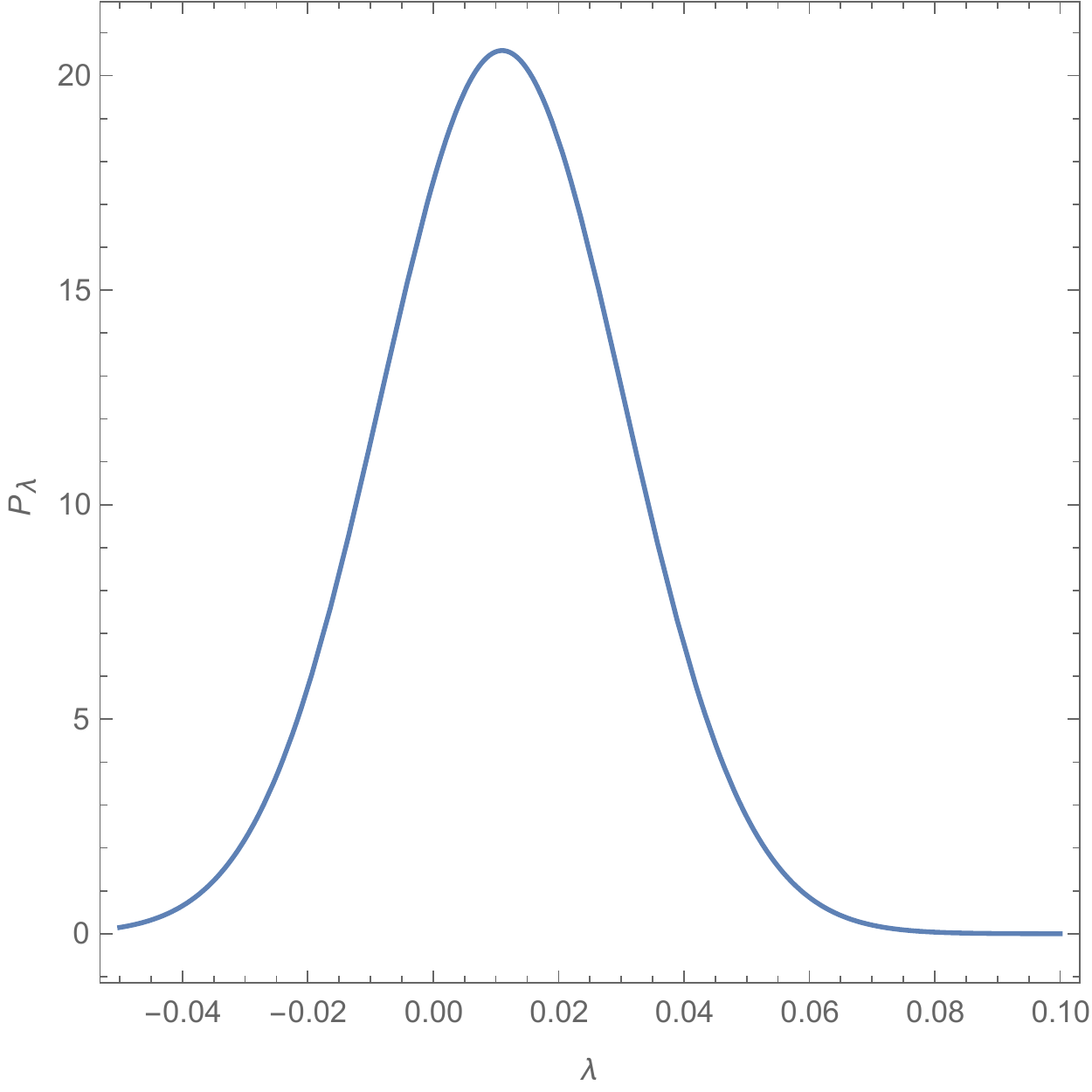}}
\end{minipage}
\caption{The first figure shows observational constraints on parameters $(\lambda,\Omega_{m0})$ with the
combination of SN Ia and BAO/CMB detasets, in which the region inside orange line and blue line are corresponding to $2\sigma$ and $1\sigma$ region respectively, while the black point (0.011,0.278)
represents the best-fit value of the pair $(\lambda,\Omega_{m0})$. The second figure is the likelihood function
of $\Omega_{m0}$ which is marginalized with a flat prior that becomes zero if $\lambda$ bigger than 0.1 or smaller
than $-0.05$ .The third figure is the likelihood function
of $\lambda$ which is marginalized with a flat prior that becomes zero if $\Omega_{m0}$ bigger than 0.4 or smaller
than 0.15.}
\label{fig:7}
\end{figure}

Now let us introduce the coupled dark energy model proposed in \cite{Yao2018}, such model can also be defined by the general field theory above with the follow two assumptions
\begin{eqnarray}\label{}
  N &=&1 \\
  I &=&m[(\frac{\kappa^{2}}{6}A^{2})]^{\frac{\lambda}{2}}=(\kappa \mid X \mid)^{\lambda}
\end{eqnarray}
Confronting such model by using the likelihood function constructed in the way in \cite{Yao2018} with new data we use above,
we have the fitting result in Fig.\ref{fig:7} and Tab.\ref{tab:2}.
In \cite{Yao2018}, we showed that
\begin{equation}
  \omega_{Xeff} \rightarrow \frac{-1+\lambda\frac{\Omega_{m0}}{1-\Omega_{m0}}(1+z)^{3(1+\lambda)}}{1+\frac{\Omega_{m0}}{1-\Omega_{m0}}\left((1+z)^{3(1+\lambda)}-(1+z)^{3}\right)}
\end{equation}
when $x_{0}\rightarrow +\infty$.
So we have $\omega_{Xeff} \approx -1+\lambda\frac{\Omega_{m0}}{1-\Omega_{m0}}-\frac{3(-2\Omega_{m0}\lambda+2\Omega_{m0}^2\lambda-\Omega_{m0}\lambda^2+2\Omega_{m0}^2\lambda^2)z}{(1-\Omega_{m0})^2}$
when redshift close to $0$. Therefore, one finds that the effective EOS of dark energy crosses $-1$ when $z=\frac{\lambda\Omega_{m0}}{3(-2\Omega_{m0}\lambda+2\Omega_{m0}^2\lambda-\Omega_{m0}\lambda^2+2\Omega_{m0}^2\lambda^2)}$, substituting in the best fitting parameters, we have $z=-0.23$. So, such coupled canonical three-form dark energy model predicts that effective EOS of dark energy crosses the cosmological constant boundary $-1$ in the not long future. The different crossing behaviors between the noncanonical model and the canonical model is due to the difference between attractors of two model. In the noncanonical model, the best-fit attractor behaves like a phantom, as a result, it ahead of time of the crossing. However, in the canonical model, the attractor is a de Sitter attractor \cite{Yao2018}, so it have little effect on the crossing moment. This result suggests that there is a possibility for us to propose a coupled noncanonical three-form dark energy model by using (1) and assumptions about $N$ and $I$, for the purpose to construct quintom scenario in the future observations.

\begin{table}
\begin{center}
\begin{tabular}{cc|  cc }
\hline\hline parameters & &  confidence interval
\\ \hline
$\Omega_{m0}$    && $ 0.278_{-0.046}^{+0.046}$
                     \\
$\lambda$         &&  $0.011_{-0.039}^{+0.039}$
                     \\
 \hline $AIC$  &&  $1042.09$
                      \\
\hline\hline
\end{tabular}
\caption{$2\sigma$ confidence interval of the model parameters and the AIC of the model.}
\label{tab:2}
\end{center}
\end{table}

\section{Conclusions}
\label{sec:4}
In this paper we have studied a coupled dark energy model which considers dark
energy as a noncanonical three-form field and dark matter as dust. By performing a dynamical analysis on the field
equations with the introduction of two dimensionless variables, we
obtained two fixed points of the autonomous system of evolution equations, one is a stable point, and the other is a tracking saddle point which provides a possible solution of the coincidence problem.

By marginalizing over $x_{0}$, we have carried out a likelihood analysis on the parameters $\lambda$ and $\Omega_{m0}$ with the combination of SN $\uppercase\expandafter{\romannumeral1}$a+BAO/CMB datasets, with the result that $\Omega_{m0}= 0.280_{-0.048}^{+0.048}$ and $\lambda=0.011_{-0.032}^{+0.032}$ in the $2\sigma$ confidence level. We also find that the best fitting effective dark energy EOS crosses $ -1$ at redshift around 0.2. As a comparison, we introduce a coupled canonical three-form dark energy model proposed in our previous work and confront it with the new data, with the results that $\Omega_{m0}= 0.278_{-0.046}^{+0.046}$, $\lambda=0.011_{-0.039}^{+0.039}$ in the $2\sigma$ confidence level. And in this model the best-fit effective dark energy EOS crosses the cosmological constant boundary $ -1$ at redshift $-0.23$. We point out that the different crossing behaviors between two model is due to the difference between two attractors. In the end, we indicate that in order to construct quintom scenario that in consistent with the future cosmological observations, one may turn to a coupled noncanonical three-form dark energy model with proper assumptions about $N$ and $I$.

In the end, we need to point out that structure formation data are necessary to obtain a more stringent constriction for the parameters, and such data also may exclude our model and the best-fit coupling parameter.

\section*{Acknowledgments}

\bibliographystyle{JHEPmodplain}
\bibliography{g4}

\end{document}